\documentclass[10pt]{article} 
\usepackage[dvips]{graphicx} 

\usepackage{hyperref}

\begin{document}


\title{Compact Quasi-Chaotic Optical Cavity}

\author{Jonathan A. Fan$^*$, Evgenii E. Narimanov, and Claire Gmachl \\
 \\
Department of Electrical Engineering, and PRISM, \\ Princeton University, Princeton, NJ 08544}

\date{ }

\maketitle

\begin{abstract}
A novel, 3-dimensional, convex, multi-pass optical cavity with partially-chaotic ray dynamics is presented. The light is localized near stable, long-path length trajectories supported by the cavity, and beam diffraction is suppressed by the phase space barriers between the regions of regular and chaotic ray dynamics that are generally present in partially-chaotic systems. For a 
centimeter-size cavity, the design supports meter-scale optical path lengths, suggesting future applications in trace gas detection.  
An exemplary cavity has been fabricated from a hollow, gold-coated, acrylic shell.   Our measurements using a HeNe laser and 
a pulsed red diode laser for characterization of the cavity beam pattern and optical path length, respectively,
confirm the theoretically predicted optical dynamics and the ability of the cavity to support meter-scale path lengths.  
\end{abstract} 


$$ { \ } $$

Optical cavities supporting long path lengths are widely used in applications such as trace gas sensing, where sensitivity strongly correlates with path length.  With optical path lengths on the order of ten meters, gas detection sensitivities on the order of parts per million or better are achieved.  Currently, the cavities most commonly used to generate long path lengths are multi-pass cells such as White Cells and Herriott Cells.\cite{Sigrist}  These devices operate by reflecting a beam between multiple mirrors over many passes, and are designed under the premise of integrable (regular) beam dynamics.  While such devices successfully provide for long optical path lengths, they have a number of drawbacks. In particular, these systems consist of multiple focusing components, take time to align properly for long path lengths, and are typically several tens of centimeters or even a meter in length.  A compact, robust, cost-effective device with straightforward optical alignment would be ideal for spectroscopy applications requiring portability and durability.  In the present Letter, we present a compact, economical, monolithic optical cavity with straightforward alignment.

The cavity design takes advantage of the partially-chaotic ray dynamics in optical systems with reduced symmetry. When the cavity shape is deformed from symmetrical geometry, the corresponding phase space generally shows a mixture of stable and chaotic 
regions.\cite{Gutzwiller}  This behavior can be accurately described using the standard methods of nonlinear 
dynamics,\cite{WGNature,Chang00,LaceyWang01}   and is the foundation of the recently demonstrated high-power semiconductor micro-cylinder 
lasers.\cite{Science98}  The barriers that separate different stable and chaotic regions in the phase space cannot be crossed by any classical (ray) trajectory - thus leading to light confinement within the phase space region into which the light beam was initially injected. While wave effects can violate such purely ``classical'' restrictions, the corresponding evanescent processes have exponentially small probability and in the case of centimeter-scale device can be essentially ignored. The advantage of using the partially-chaotic cavity is thus two-fold. First, it will assure light confinement near a stable trajectory if light is 
injected into the cavity's 
region of stability. Second, the resulting inaccessibility of the chaotic regions to the injected light allows efficient gas input/output for the cavity if the corresponding gas ports are placed in such inaccessible regions.

\begin{figure}[tb]
\centerline{\includegraphics[width=7.cm]{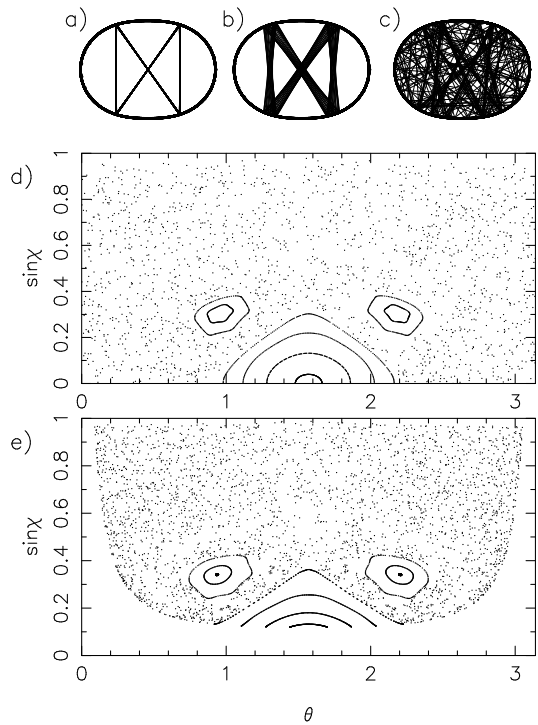}}
\caption{
\label{fig:SOS}
Top panel: ray dynamics in a quadrupole with $\epsilon = 0.16$ for injection into the bow-tie orbit (a), its stability island (b) and into chaotic phase space (c).  Panels (d) and (e) show the Poincar\'e Surface of Section (SOS), for angular momentum $L_z = 0$ (d)
and $L_z=0.15$ (e).   Note that the chain of two bow-tie islands is present in both (d) and (e). 
}
\end{figure}

While various topologies of the desired stable multi-pass trajectory are possible, the optimal design corresponds to the orbit that has the same angle of incidence at all reflection points - as this would allow an efficient use of ultra-high reflection interference coatings, thus minimizing the energy loss on every reflection. In particular, this requirement is satisfied by the `"bow-tie" orbit shown in Fig. \ref{fig:SOS}(a, b). In terms of practical issues of manufacturing the cavity, the desired geometry still has cylinder symmetry - as it would allow the fabrication of the device using precision diamond turning.

\begin{figure}[tb]
\centerline{\includegraphics[width=7.5cm]{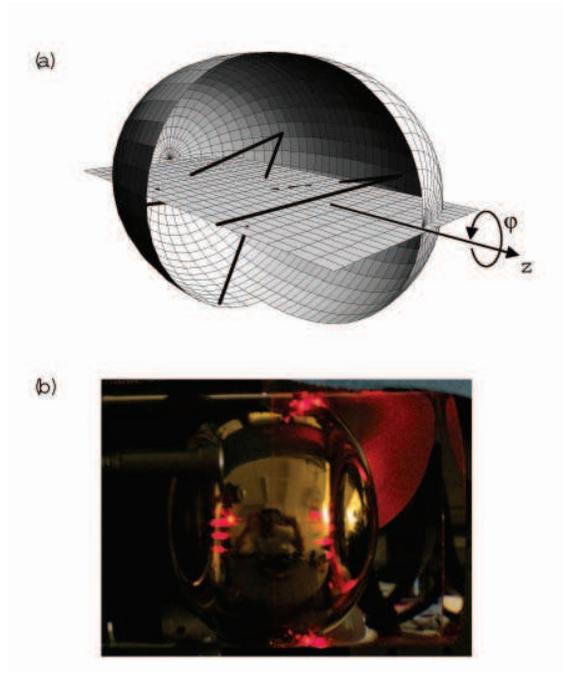}}
\caption{
\label{fig:cavity}
(a)  
Cutaway view of the 3D cavity with a single offset "bow-tie" pass visualized.  
The plane is a guide to the eye; the beam is injected into the cavity at a small angle to this plane.
(b) Photograph of the cavity under illumination with the HeNe laser. The average radius of the cavity is 2.54 cm. The ray dynamics in the cavity can be characterized by the size and position of the spots created from transmitted HeNe laser light at each point of reflection in the cavity.   
}
\end{figure}

The geometry we have therefore chosen for our compact, quasi-chaotic, multi-pass cavity, is the 
three-dimensional (3D) quadrupole, defined in terms of its average
radius $R_0$ and the deformation parameter $\epsilon$ as
$
R  =  R_0 \left(1 + \epsilon \cos 2\theta\right)
$
in the standard spherical coordinates $(R,\theta,\phi)$. It can be
visualized as an oval rotated  around its major axis $z$ - see Fig. \ref{fig:cavity}(a). The resulting ray dynamics for this cavity is partially-chaotic -- as demonstrated in the phase space plots (also known as ``Poincar\'e Surfaces of Section (SOS)) in Fig. \ref{fig:SOS}(d,e) that visualize the evolution of the  ensemble of rays of varying input angles and positions.   As clearly seen from Fig. 
\ref{fig:SOS} (d,e), the proposed design supports the stable bow-tie orbit - its stability region is clearly seen as the ``chain" of two ``islands" centered at the same value of $\sin\chi$ (where $\chi$ is the angle of incidence). This (three-dimensional) bow-tie trajectory can be visualized by offsetting the ``2D" bow-tie above the 2D quadrupole plane, in a manner indicated 
in Fig. \ref{fig:cavity}(a).

Due to the confinement of the trajectories to the corresponding (stable) region of the phase space, the injected rays whose angle of incidence and position reside in one of the islands remain coupled in the stable mode over the course of multiple reflections - oscillating near the  bow-tie orbit (see Fig. \ref{fig:SOS} (b)) while the plane that contains the orbit rotates around the axis of symmetry of the device (with the angular velocity determined by the angular 
momentum $L_z$ with respect to the axis of symmetry $z$.  Rays injected at initial parameters outside the stable islands, {follow chaotic trajectories}, and eventually fill the entire cavity, as shown in Fig. \ref{fig:SOS}(c).

With a single aperture in the cavity positioned at a reflection point of the bow-tie trajectory drawn in Fig. \ref{fig:SOS}(a), light can be coupled into the cavity with a small but nonzero value 
of $L_z$ , undergo multiple passes around the cavity's axis of cylindrical symmetry, and be coupled out for detection at an angle different from the input angle.

Due to the axial symmetry of the proposed device, the dynamics in the corresponding dimension (given by the polar angle $\phi$ - see 
Fig. \ref{fig:cavity}(a)) is marginally stable (or metastable). ); i.e. for the rays injected with different values of the angular 
momentum  $L_z$, the deviations growth linearly with the number of bounces (as opposed to exponential growth for unstable/chaotic dynamics). This implies that  the beam spot size -- while being {\it limited by the size of the island} in the  the  $\theta$-direction, -- growth linearly with time in the $\phi$-direction.\cite{sphere} While this metastable condition allows for beam divergence over the course of multiple reflections, long path lengths can still be supported if the linear growth factor is small and only present in one direction, minimizing divergence (to the limit set by the beam diffraction).

An experimental cavity has been constructed to test theoretical stability projections and multiple-pass dynamics, 
with $R_0 = 2.54$ cm, and $\epsilon = 0.16$.  
The cavity comprises two halves of an acrylic plastic shell carved out by diamond-turning , and gold is uniformly deposited inside 
the cavity.\cite{manufacturer}  A circular aperture $2$ mm in diameter is drilled along the line $\phi = 54.4^\circ$ at one point in the cavity.  
Light from a $15$ mW HeNe laser or red laser diode is coupled into the cavity using a spherical lens with a focal point of $50.8$ mm mounted on an 
$x-y-z$ translation stage, and the precision of the stage allows to inject the 
light into an island of stability.  
Light coupled out of the cavity is focused through a pair of spherical lenses, detected by a high-speed Si detector, and analyzed using a digital oscilloscope.

	The advantage of using a visible laser is that the light is partially transmitted through the gold coating, 
so that the beam at points of 
reflection is visible from the outside of the cavity (See Fig. \ref{fig:cavity}(b)).  Thus, the beam shape as a function of 
the number of reflections can be assessed and used to characterize the cavity's stability.  By correlating the experimental spot position (Fig. \ref{fig:cavity}(b)) with multiple-pass ray dynamics from Fig. \ref{fig:SOS}, the number of reflections the beam has undergone before reflecting off a particular position can be determined.  Measurements indicate that the spot size in the quadrupole plane grows quickly to a constant width over the course of multiple reflections, confirming stability in the bowtie mode.  In the direction perpendicular to the quadrupole plane, spot size is measured to increase linearly as a function of number of reflections, confirming marginal stability in the system for rotation around the cavity's axis of cylindrical symmetry.

\begin{figure}[tb]
\centerline{\includegraphics[width=8.cm]{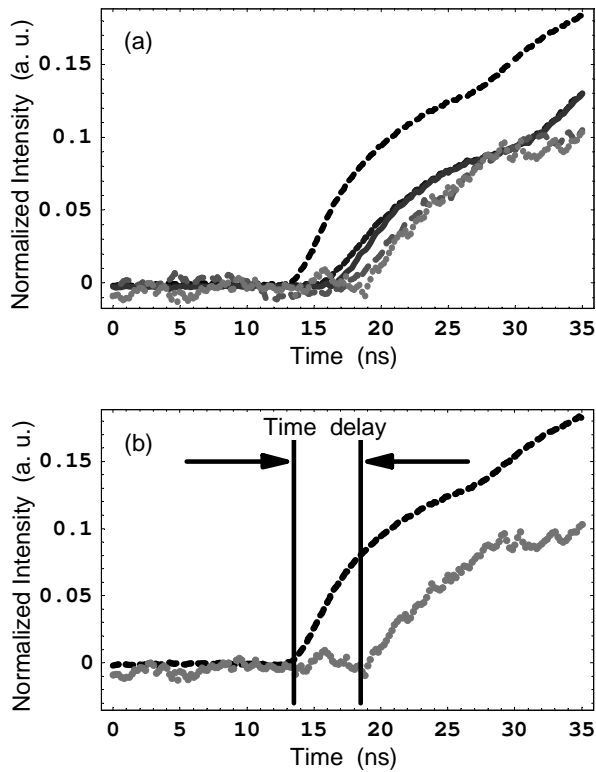}}
\caption{
\label{fig:data}
(a)  Plot of the leading edges of five detected pulses with varying time delays corresponding to (left-to-right) 1, 4, 5, 7, and 8 passes through the cavity.   (b)  Pulse time delay data for 8 pass (dotted) and single pass (dashed) alignment:
 a difference of 5.0 ns (~152 cm) between the multiple pass and single pass signal is measured.
}
\end{figure}

	Path lengths in the cavity can be measured by a pulse time delay setup.  The experimental concept involves coupling short pulses of light with well-defined temporal envelopes into the cavity.  First, the cavity is oriented to allow just a single pass in the quadrupole plane before light is coupled out of the resonator and detected.  Then the cavity is offset from the quadrupole plane to allow an input pulse to undergo multiple bow-tie passes before coupling out and being detected.  By measuring the time delay between the multiple pass and single pass pulse edges, the path length difference is measured.  In the experiment, an AlGaInP diode laser with $30$ mW maximum power is used and pulsed with an HP pulse generator.  The temporal pulse widths are 40 ns, the duty cycle is $0.01$\%, and the pulse edge rise times are on the order of nanoseconds.  Noise in the data is reduced by averaging over 512 consecutive pulses by use of a digital oscilloscope, and the background signal obtained without light pulses is subtracted.  Normalized time-resolved pulse data for five different multiple-pass schemes is given 
in Fig. \ref{fig:data}(a).  From left to right, the pulse edges are for multiple-pass schemes of $1$, $4$, $5$, $7$, and $8$ passes.  
 (b) shows the time delay between 1 and 8 passes.  The measured time-delay is 5.0 ns, which corresponds to ~152 cm of path length in addition to a single 2D quadrupole pass.  Given that the single bow-tie mode in the 2D quadrupole plane measures $17.2$ cm, the total path length in the orbit is $\sim 169$ cm.

Run-time measurements for longer path lengths, while possible, were difficult due to the gold transmission and absorption losses and scattering effects due to surface roughness in the cavity. In future, this issue can be addressed by use of ultra-high-reflectivity interference coatings instead of the partially transmitting gold film, as well as by using longer wavelengths. 

To summarize, we presented a novel approach to compact multi-pass optical cavities based on partially-chaotic ray dynamics. The stability of the spot size of a properly injected beam over multiple reflections was confirmed experimentally with a custom-built cavity, and path lengths on the order of a meter were measured.
 
\noindent 

This work was partially supported by NSF grants DMR-0134736, ECS-0400615, and the Princeton Institute for the Science and Technology of Materials (PRISM). 

\noindent$^*$Present address: Harvard University, Cambridge MA 02138; jfan@fas.harvard.edu



\begin{thebibliography}{99}


\bibitem{Sigrist} M.~Sigrist, Air Monitoring by Spectroscopic Techniques (John Wiley \& Sons, New York, 1994).

\bibitem{Gutzwiller} M.~Gutzwiller, Chaos in Classical and Quantum Mechanics (Springer-Verlag, New York, 1991).

\bibitem{WGNature}J.~U.~N\"{o}ckel, A.~D.~Stone, {\it Ray and wave chaos in asymmetric resonant optical cavities}, Nature {\bf 385}, 45 (1997)

\bibitem{Chang00}
S.~Chang, R.~Chang, A.~D.~Stone, and J.~Nockel, ``Observation of Emission from Chaotic Lasing Modes in Deformed Microspheres: Displacement by the Stable-Orbit Modes," J. Opt. Soc. Am. B 17, 1828-1834 (2000) 

\bibitem{LaceyWang01}
S. Lacey and H. Wang, ``Directional Emission from Whispering-Gallery Modes in Deformed Fused-Silica Microspheres," Opt. Lett. 26, 1943-1945 (2001).

\bibitem{Science98}
C.~Gmachl, F.~Capasso, E.~E.~Narimanov, J.~U.~Nockel, A.~D.~Stone, J.~Faist, D.~L.~Sivco, and A.~Y.~Cho, ``High-Power Directional Emission from Microlasers with Chaotic Resonators," Science 280, 1556-1564 (1998).
\bibitem{manufacturer} http://www.syntectechnologies.com/index.htm

\bibitem{sphere} 
Note that in a sphere the size of the beam spot will grow (linearly) in {\it both} directions.

\end{thebibliography}
\end{document}